\documentclass[twocolumn,showpacs,preprintnumbers,amsmath,amssymb]{revtex4}
\usepackage{dcolumn}
\usepackage{bm}
\usepackage{graphicx,subfigure,xcolor}

\newcommand{\bx}{\textbf{x}}
\newcommand{\bxp}{\textbf{x}'}

\newcommand{\adl}{a^{\dag}_\ell}
\newcommand{\al}{a_\ell}
\newcommand{\adj}{a^{\dag}_{j}}
\newcommand{\aj}{a_{j}}

\newcommand{\Cjl}{C_{j\ell}}

\newcommand{\Djl}{D_{j\ell}}
\newcommand{\CkjG}{C_{\textbf{kj}}}
\newcommand{\DkjG}{D_{\textbf{kj}}}

\newcommand{\wosc}{\omega_{osc}}

\newcommand{\wj}{\omega_{j}}
\newcommand{\wl}{\omega_{\ell}}

\newcommand{\CCp}{\frac{\hbar c^2}{L_0\omega_p}}

\newcommand{\CCpL}{\frac{\hbar c^2}{2\omega_p}}

\newcommand{\adkG}{a^{\dag}_{\textbf{k}}}
\newcommand{\akG}{a_{\textbf{k}}}
\newcommand{\adjG}{a^{\dag}_{\textbf{j}}}
\newcommand{\ajG}{a_{\textbf{j}}}

\newcommand{\urG}{u_{\textbf{r}}}

\newcommand{\adkGl}{a^{\dag}_{\textbf{k}0}}
\newcommand{\akGl}{a_{\textbf{k}0}}
\newcommand{\adjGl}{a^{\dag}_{\textbf{j}0}}
\newcommand{\ajGl}{a_{\textbf{j}0}}
\newcommand{\wkG}{\omega_{\textbf{k}}}
\newcommand{\wjG}{\omega_{\textbf{j}}}
\newcommand{\wpG}{\omega_{\textbf{p}}}

\newcommand{\wrG}{\omega_{\textbf{r}}}
\newcommand{\wkGl}{\omega_{\textbf{k}0}}
\newcommand{\wjGl}{\omega_{\textbf{j}0}}

\begin{document}

\title{Vacuum energy densities of a field in a cavity with a mobile boundary}

\author{Federico Armata}
\altaffiliation[Present address: ]{QOLS, Blackett Laboratory, Imperial College London, London SW7 2BW, United Kingdom
}
\email{f.armata@imperial.ac.uk}

\author{Roberto Passante}
\email{roberto.passante@unipa.it}

\affiliation{Dipartimento di Fisica e Chimica, Universit\`{a} degli Studi di Palermo and CNISM, Via Archirafi 36, I-90123 Palermo, Italy}


\begin{abstract}
We consider the zero-point field fluctuations, and the related field energy densities, inside a one-dimensional and a three-dimensional cavity with a mobile wall. The mechanical degrees of freedom of the mobile wall are described quantum mechanically and they are fully included in the overall system dynamics. In this optomechanical system, the field and the wall can interact with each other through the radiation pressure on the wall, given by the photons inside the cavity or even by vacuum fluctuations. We consider two cases: the one-dimensional electromagnetic field and the three-dimensional scalar field, and use the Green's functions formalism, which allows extension of the results obtained for the scalar field to the electromagnetic field. We show that the quantum fluctuations of the position of the cavity's mobile wall significantly affect the field energy density inside the cavity, in particular at the very proximity of the mobile wall. The dependence of this effect from the ultraviolet cutoff frequency, related to the plasma frequency of the cavity walls, is discussed. We also compare our new results for the one-dimensional electromagnetic field and the three-dimensional massless scalar field to results recently obtained for the one-dimensional massless scalar field. We show that the presence of a mobile wall also changes the Casimir-Polder force on a polarizable body placed inside the cavity, giving the possibility to detect experimentally the new effects we have considered.
\end{abstract}

\maketitle

\section{\label{sec:level1}Introduction}

The existence of vacuum field fluctuations, and the related vacuum energy density of the field, is a striking consequence of quantum electrodynamics and quantum field theory in general \cite{Milton01}. Vacuum fluctuations have observable effects, for example the Casimir force, which is usually an attractive force of quantum origin between neutral macroscopic metallic or dielectric bodies placed in the vacuum space \cite{Casimir48,Milonni94}. Casimir forces originate from the change of the field energy associated with the vacuum fluctuations, when one or more boundary conditions such as dielectric or conducting objects, are changed.

A more thoroughly description of quantum vacuum effects can be obtained by considering local field quantities such as the field energy density. Obtaining the energy densities of the field in the vacuum state is relevant also in view of the fact that they are strictly related to atom-surface Casimir-Polder dispersion interactions \cite{PPT98,Buhmann12}. Moreover, it has been shown that the vacuum electric and magnetic energy density, as well as vacuum field fluctuations, can become singular in the proximity of sharp metallic boundaries \cite{SF02,MCPW06,BP12}. The presence of such surface divergences of the energy density could be relevant also in view of the fact that they should yield significant gravitational effects because the energy density acts as a source term for gravity \cite{Milton11,MNS11}.

New effects arise when one or more boundaries are allowed to move. A known effect is the dynamical Casimir effect, that is the emission of real quanta from the vacuum when a boundary is forced to move with nonuniform acceleration \cite{Moore70,Law94, Dodonov10}. For a very small mass of a mobile wall, quantum effects relative to its mechanical degrees of freedom such as position fluctuations, may be relevant. It is thus worth to consider the effect of a mobile boundary, whose mechanical degrees of freedom are treated quantum mechanically, on a quantum field; in this case an effective coupling between the wall and the field arises due to the radiation pressure, as well as an effective coupling between the field modes \cite{LawField-Mirror}. These effects are also related to the growing field of quantum optomechanics, which studies the coupling of optical cavity modes with mechanical degrees of freedom \cite{MG09,Meystre13,AKM13}; this subject is also relevant for building more sensitive force detectors, to be used for example for the detection of gravitational waves \cite{KV08}. The effect of vacuum fields on the position fluctuations of a single mirror in the vacuum space, and the role of vacuum friction, has been recently considered \cite{WU14}.

We consider in this paper how a moving conducting boundary such as a cavity wall, being treated quantum mechanically, can affect the field fluctuations and the related field energy density inside the cavity. This effect can be in principle observed because the field energy density can be probed through the Casimir-Polder interaction with a polarizable body placed inside the cavity. In a previous paper \cite{Pass-Butera}, we have investigated this aspect in the simple case of a massless scalar field in a one-dimensional cavity with one fixed and one mobile wall, and we have found a change of the field fluctuations in the cavity, particularly relevant in the proximity of the mobile wall, and of the Casimir force between the cavity walls. Also, this new effect, in the case considered in \cite{Pass-Butera}, has a size such that it should allow its experimental observation for a sufficiently small mass of the mobile wall, and masses down to $10^{-21}$ Kg can be nowadays reached in modern optomechanics experiments \cite{AKM13}.

In this paper we extend the results previously obtained for the one-dimensional massless scalar field to the more realistic cases of a one-dimensional electromagnetic field and the three-dimensional massless scalar field. We give a local description of vacuum field fluctuations in terms of the zero-point field energy density and local field fluctuations, inside a cavity with a mobile wall, both for the one-dimensional electromagnetic field and the three-dimensional scalar field. In both cases the motion of the mobile wall, which is assumed bound to an equilibrium position by a harmonic potential, is described quantum-mechanically and the effects of radiation pressure and of the wall's quantum position fluctuations are included in the formalism. Our description is based on an appropriate generalization, that we introduce in this paper, of the effective Hamiltonian obtained in \cite{LawField-Mirror} for the one-dimensional massless scalar case. For the one-dimensional electromagnetic case, using a formalism based on the Green's functions, we obtain the change of the renormalized electric and magnetic energy densities inside the one-dimensional cavity due to the motion of the cavity wall; we show that, similarly to the one-dimensional scalar results, it is particularly relevant close to the equilibrium position of the mobile wall. We also discuss their dependence on the ultraviolet cutoff frequency, that is related to the plasma frequency of the mobile wall, and show that, when the cutoff frequency is increased, the energy-density change becomes more and more concentrated near the mobile wall. In the three-dimensional scalar case, we first obtain the renormalized Green's function of the scalar field on the interacting ground state, and we then use it to obtain the renormalized field energy density change in the cavity. We show that, contrarily to both scalar and electromagnetic one-dimensional cases, the peak of the energy density change is not located at the equilibrium position of the mobile wall. This peak, however, moves towards the mobile wall when the cutoff frequency is increased. We also discuss the dependence of the change of the field energy density on the mass and oscillation frequency of the mobile wall, and how the new effects we have obtained can be observed through the Casimir-Polder force on a polarizable body placed inside the cavity in the proximity of the mobile wall.

Although our system has some analogy with the dynamical Casimir effect, we wish to stress that it is quite different, because in the present case the mobile wall is not moving according to a prescribed law as in the dynamical Casimir effect, but it is described quantum mechanically according to the quantum dynamics induced by the Hamiltonian of the interacting wall-field system. Finally, we point out that the results we obtain for the three-dimensional scalar field can be also useful for an extension to the three-dimensional electromagnetic field case.

This paper is organized as follows. In Sec. \ref{sec:local-formalism} we introduce part of the local formalism that will be used in the subsequent sections of the paper.
In Sec. \ref{sec:level2} we consider the problem of the interaction between a one-dimensional electromagnetic field and a movable wall, using the local formalism and exploiting the results obtained in Refs. \cite{LawField-Mirror, Pass-Butera} for the simpler case of a massless scalar field in a one-dimensional cavity. We obtain the change of the renormalized zero-point energy densities of the electric and magnetic field components on the interacting ground state of the system. In Sec. \ref{sec:level3} we consider the case of a massless scalar field in a three-dimensional cavity with one mobile wall. We first generalize the Law's effective Hamiltonian to the three-dimensional case, and then obtain the correction to the field energy density consequent to the wall's motion due to quantum position fluctuations and radiation pressure.
We discuss the main physical features of the change of the field energy density and compare our new results with previous ones obtained for the one-dimensional scalar case and discuss observability of the new effects found. Section \ref{sec:Conclusion} is devoted to our conclusive remarks.

\section{\label{sec:local-formalism}The local formalism}
In this section we introduce the local field formalism we will use in the following and the relative notations. A field theoretical approach to the study of the properties of the vacuum starts from the analysis of the behavior of local field quantities. For our purposes, the energy-momentum tensor $T^{\mu\nu}$ represents the appropriate quantity, because $T^{00}$ is the energy density of the field, the components $T^{0\nu}$ are related to the energy and momentum flow, and the stress components $T^{ik}$ are related to general mechanical properties of the vacuum.

In the presence of boundaries, a local formulation requires the introduction of the (renormalized) energy-momentum tensor of the vacuum $\Theta^{\mu\nu}_{vac}$, in the form \cite{Brown-Maclay,Candelas,Greiner}
\begin{equation}\label{Teta-Vacuum}
  \Theta^{\mu\nu}_{vac}=\langle0|T^{\mu\nu}|0\rangle_{\partial\Gamma}-\langle0|T^{\mu\nu}|0\rangle_0 \, .
\end{equation}

In this equation, the measurable vacuum energy-momentum tensor is defined as the difference between that in the confined field configuration $\langle0|T^{\mu\nu}|0\rangle_{\partial\Gamma}$ and that corresponding to the unbounded configuration $\langle0|T^{\mu\nu}|0\rangle_0$. An advantage of a local description is that it permits a different and thoughtful point of view yielding a deeper understanding of the nature of vacuum energy and vacuum stresses. It is known that $\Theta^{\mu\nu}_{vac}$ can be expressed in terms of the field propagators. The presence of quantum field fluctuations in a specific configuration, and the consequent observable quantum vacuum effects, can be understood from the modifications of the emission/reabsorption of virtual field quanta under external constraints. When boundaries are introduced, the propagation is modified due to surface interactions, and consequently this perturbs the homogeneity of the corresponding propagator.
For example, the propagator of the scalar field $G(x,x')$ must satisfy the appropriate boundary conditions, for instance Dirichlet or von Neumann boundary conditions in the case of perfect reflectors. The energy-momentum tensor of the electromagnetic vacuum for the confined field configuration, can be written as \cite{Candelas,Greiner}
\begin{equation}\label{Teta-vuoto-BC}\begin{split}
  \Theta^{\mu\nu}_{vac}(x)&=-i\left\{\tau^{\mu\nu}_{x,x'}\left(G(x,x')-G_0(x-x')\right)\right\}\Big|_{x'=x}\\
&=-i\left\{\tau^{\mu\nu}_{x,x'}G_R(x,x')\right\}\Big|_{x'=x} \, ,
\end{split}\end{equation}
where $G$ is the scalar propagator in the presence of the boundaries, $G_0$ is the scalar propagator  in the unbounded space, and $G_R=G-G_0$ is the renormalized propagator. We have also introduced the differential operator
\begin{equation}\label{tau}
  \tau^{\mu\nu}_{x,x'}=2\left(\partial^{\mu}\partial'^{\nu}+\frac{1}{4}g^{\mu\nu}\partial^{\alpha}\partial'_{\alpha}\right) \, ,
\end{equation}
with $g^{\mu\nu}=\text{diag(-1,+1,+1,+1)}$. The vacuum subtraction in Eq. $\eqref{Teta-Vacuum}$ is now obtained from the difference between the confined and the free propagators, respectively, $G(x,x')$ and $G_0(x-x')$, which is the renormalized Green function of the system $G_R(x,x')$. For explicit evaluations, all one has to do is construct $G_R(x,x')$ for the considered configuration (see Refs. \cite{Brown-Maclay,Candelas,Greiner,Lukosz,Lukosz2,Ford-Svaiter}, for example) that, except for special cases with a simple geometry, can be a challenging task.
The local method has, however, the advantage that it allows us to obtain local quantities of the electromagnetic field, such as its energy density, from the Green's functions of the scalar field with Dirichlet and von Neumann boundary conditions relative to the problem under investigation, using Eq. \eqref{Teta-vuoto-BC} \cite{Lukosz, Lukosz2}. We will use this formalism for tackling our problem in the next section. In addition, electric and magnetic field two-point correlation functions can be also obtained from the scalar propagator, using \cite{Brown-Maclay,Ford-Svaiter}
\begin{eqnarray}\label{D-muni}
D^{\mu\nu;\lambda\kappa}(x-x')&=&i\langle 0 \mid F^{\mu \nu}(x) F^{\lambda \kappa}(x') \mid 0 \rangle
\nonumber \\
&=& d^{\mu\nu;\lambda\kappa} G_0(x-x') \, ,
\end{eqnarray}
where $F^{\mu \nu}$ is the electromagnetic strength tensor and we have defined the differential operator
\begin{equation}\label{d-muni}
d^{\mu\nu;\lambda\kappa}=\partial^{\mu}\partial'^{\lambda}g^{\nu\kappa}-\partial^{\nu}\partial'^{\lambda}g^{\mu\kappa}+\partial^{\nu}\partial'^{\kappa}g^{\mu\lambda}
-\partial^{\mu}\partial'^{\kappa}g^{\nu\lambda} \, .
\end{equation}
In the next sections we will use Eq. $(\ref{D-muni})$, by applying the operator $(\ref{d-muni})$ to the renormalized Green's function in the presence of the boundaries.

\section{\label{sec:level2}The one-dimensional electromagnetic case}

In order to consider the electromagnetic field inside a one-dimensional cavity using the approach outlined in the previous section, we first consider a one-dimensional cavity formed by two perfectly reflecting mirrors and a massless scalar field $\phi (x,t)$ at zero temperature. One of the mirrors is fixed at the position $x=0$ while the other is bounded by a harmonic potential $V(q)$ to its equilibrium position $L_0$, and has mass $M$ and oscillation frequency $\wosc$. We label the position  of the movable mirror by $q(t)$, which is an operator because we are treating the mirror's motion quantum-mechanically. The effective nonrelativistic Hamiltonian describing our one-dimensional coupled mirror-field system is $H=H_0+H_{int}$, where
\begin{equation}\label{H0}
  H_{0}=\hbar\wosc b^{\dag}b+\hbar\sum_{j}\omega_j\adj\aj
\end{equation}
is the unperturbed Hamiltonian. The first and second term of \eqref{H0} are, respectively, the mirror and the field Hamiltonian, with: $b$ and $b^\dagger$ annihilation and creation operators of the mechanical degrees of freedom of the movable mirror;  $\aj$ and $\adj$ annihilation and creation operators for the mode $j$ of the scalar field.
We impose Dirichlet boundary conditions on the field operator;
the field modes are relative to the equilibrium position $L_0$ of the moving mirror, and thus the possible wave numbers are $k_j=j\pi /L_0$, with $j$ an integer number.
The effective interaction Hamiltonian, describing the mobile mirror-field interaction and an effective interaction between different field modes (due to the motion of the wall), is \cite{LawField-Mirror}
\begin{equation}\label{Hint-Giulio}
  H_{int}=-\sum_{j\ell}\Cjl{(b+b^{\dag})\mathcal{N}[(\aj+\adj)(\al+\adl)]} \, ,
\end{equation}
where
\begin{equation}\label{Ckj}
  \Cjl=(-1)^{j+\ell}\left(\frac{\hbar}{2}\right)^{3/2}\frac{1}{L_{0}\sqrt{M}}\sqrt{\frac{\wj\wl}{\wosc}}
\end{equation}
is the coupling constant and $\mathcal{N}$ is the normal ordering operator, while $j$ and $\ell$ are integer numbers specifying the field modes (evaluated for the equilibrium position of the wall).

From the Hamiltonian $(\ref{Hint-Giulio})$, using perturbation theory at the lowest significant order, it is possible to obtain the dressed ground state $|g\rangle$ of the field-mirror system as done in \cite{Pass-Butera},
\begin{equation}\label{ground-Giulio}
  |g\rangle=|\{0_{p}\},0\rangle+\sum_{j\ell}\Djl |\{1_{j},1_{\ell}\},1\rangle \, ,
\end{equation}
where the elements of the states in curly brackets indicate field excitations, while the other element indicates excitations of the wall's mechanical degrees of freedom. We have also defined
\begin{equation}\label{Dkj}
  \Djl=(-1)^{j+\ell}\frac{1}{L_{0}}\sqrt{\frac{\hbar\wj\wl}{8M\wosc}}\frac{1}{(\wosc+\wj+\wl)}
\end{equation}
In order to obtain local quantities of the field, as outlined in Sec. \ref{sec:local-formalism}, it is useful to calculate first the renormalized scalar field propagator on the dressed vacuum state $\eqref{ground-Giulio}$, which is the difference of the Green's function for the confined field with Dirichlet boundary conditions and for the free field (from now on we explicitly write space and time components),
\begin{align}\label{GR-Rin-1Da}\begin{split}
G_R(x,t;x',t')&=\langle g|\phi_{BC}(x,t)\phi_{BC}(x',t')|g\rangle \\
&-\langle\{0_r\}|\phi_{un}(x,t)\phi_{un}(x',t')|\{0_r\}\rangle \, ,
\end{split}\end{align}
where $\phi_{BC}(x,t)$ satisfies the Dirichlet boundary condition at the wall's position and $\phi_{un}(x,t)$ is the free-field operator in the unbounded space. We obtain
\begin{align}\label{GR-Rin-1D}\begin{split}
G_R(x,t;x',t')=G_{R0}(x,t;x',t')+\Delta G_R(x,t;x',t') \, ,
\end{split}\end{align}
with
\begin{widetext}
\begin{align}\label{GR0-deltaGR}\begin{split}
G_{R0}(x,t;x',t')&=\langle\{0_r\}|\phi_{BC}(x,t)\phi_{BC}(x',t')|\{0_r\}\rangle-\langle\{0_r\}|\phi_{un}(x,t)\phi_{un}(x',t')|\{0_r\}\rangle\\
&=\left(\sum_p\CCp e^{-i\omega_p(t-t')}\sin(k_px)\sin(k_px')-\int\frac{dp}{2\pi}\CCpL e^{-i\omega_p(t-t')}e^{ik_p(x-x')}\right) \, ,\\
\Delta G_R(x,t;x',t')&=8\sum_{j\ell}\sum_{r}
\frac {\hbar c^2}{L_0 (\wj \omega_r )^{1/2}}D_{j\ell}D_{\ell r}
\left[\cos(\omega_jt-\omega_rt')\right]\left[\sin(k_j x)\sin(k_rx')\right] \, .
\end{split}\end{align}
\end{widetext}

Equation \eqref{GR0-deltaGR} shows that the renormalized field propagator is given by two terms. The first term $G_{R0}$, at the zeroth order in the atom-mirror coupling, takes into account that the field is confined in the cavity and it is the difference between the fixed-wall propagator and the free propagator. The second term $\Delta G_R$ is a correction term taking into account the effective interaction between the field and the mobile mirror, and it is related to the quantum fluctuations of the position of the mobile wall.

We can now face the one-dimensional electromagnetic case. Starting from the scalar Green's function $\eqref{GR-Rin-1D}$ and \eqref{GR0-deltaGR}, using Eq. $\eqref{D-muni}$ by applying the appropriate differential operators, we can obtain the field fluctuations and energy densities (they just differ by a multiplicative factor) associated with the electric and magnetic field components along the $z$ and $y$ directions respectively. They are given by a zeroth-order term (the same obtained for fixed walls) and a first-order term, coming from the zeroth- and first-order Green's functions $\eqref{GR0-deltaGR}$, respectively. The zeroth-order terms are given by
\begin{align}\label{FluttEz2-0}\begin{split}
\langle E_z^2(x)\rangle_0&=\lim_{(x',t')\rightarrow(x,t)}c^{-2}\partial_t\partial_{t'}\langle G_{R0}(x,t;x',t')\rangle\\
&=-\frac{\hbar c\pi}{24L_0^2}-\frac{c\pi\hbar}{2L_0^2}\frac{e^{\frac{2i\pi x}{L_0}}}{\left(e^{\frac{2i\pi x}{L_0}}-1\right)^2} \, ,
\end{split}\end{align}
\begin{align}\label{FluttBy-0}\begin{split}
\langle B_y^2(x)\rangle_0&=\lim_{(x',t')\rightarrow(x,t)}\partial_x\partial_{x'}\langle G_{R0}(x,t;x',t')\rangle \\
&=-\frac{\hbar c\pi}{24L_0^2}+\frac{\hbar c\pi}{2L_0^2}\frac{e^{\frac{2i\pi x}{L_0}}}{\left(e^{\frac{2i\pi x}{L_0}}-1\right)^2} \, .
\end{split}\end{align}

Expanding these expressions in the proximity of the movable wall position $(x \simeq L_0)$, we have
\begin{equation}\label{SviluppoEz-0}
  \langle E_z^2(x)\rangle_0\simeq\frac{\hbar c}{8\pi(x-L)^2} \, ,
\end{equation}
\begin{equation}\label{SviluppoBy-0}
  \langle B_y^2(x)\rangle_0\simeq-\frac{\hbar c\pi}{12L_0^2}-\frac{\hbar c}{8\pi(x-L)^2} \, .
\end{equation}

We wish to point out that the average quadratic values of the fields have the expected divergence at the wall's (average) position \cite{BP12}.
Summing up the two terms in $\eqref{FluttEz2-0}$ and $\eqref{FluttBy-0}$, we obtain the total Casimir energy density
$(\langle E_z^2(x)\rangle_0+\langle B_y^2(x)\rangle_0)/2=-\frac{\hbar c\pi}{24L_0^2}$ inside the one-dimensional cavity with fixed walls.

We can now evaluate the first-order correction to the electric and magnetic field fluctuations, using the Green's function correction
$\Delta G_R$ in Eq. \eqref{GR0-deltaGR}. We obtain
\begin{widetext}
\begin{align}\label{FluttEz2-1}\begin{split}
\langle E_z^2(x)\rangle_1&=\lim_{(x',t')\rightarrow(x,t)}c^{-2}\partial_t\partial_{t'}\langle\Delta G_R(x,t;x',t')\rangle\\
&=\sum_{j\ell r}(-1)^{\ell +r}\frac{\hbar^2}{L_0^3M\wosc}\frac{\omega_j\omega_\ell\omega_r}{(\wosc+\omega_j+\omega_\ell )(\wosc+\omega_j+\omega_r)}\sin(k_\ell x)\sin(k_rx) \, ,
\end{split}\end{align}
\begin{align}\label{FluttBy2-1}\begin{split}
\langle B_y^2(x)\rangle_1&=\lim_{(x',t')\rightarrow(x,t)}\partial_x\partial_{x'}\langle\Delta G_R(x,t;x',t')\rangle \\
&=\sum_{j\ell r}(-1)^{\ell +r}\frac{\hbar^2}{L_0^3M\wosc}\frac{\omega_j\omega_\ell\omega_r}{(\wosc+\omega_j+\omega_\ell )(\wosc+\omega_j+\omega_r)}\cos(k_\ell x)\cos(k_rx) \, .
\end{split}\end{align}
\end{widetext}

The corrections \eqref{FluttEz2-1} and \eqref{FluttBy2-1} to the electric and magnetic energy densities take into account of the effective field-mirror interaction and of the effects of radiation pressure, related to the wall's quantum fluctuations of its position. Summing up these two corrections, we obtain the correction to the field energy density in the cavity
\begin{widetext}
\begin{equation}\label{EnergyDensity-1D-TOT}
\frac 12 \left[ \langle E_z^2(x)\rangle_1 + \langle B_y^2(x)
\rangle_1 \right] =\sum_{j\ell r}(-1)^{\ell +r}\frac{\hbar^2}{2L_0^3M\wosc}\frac{\omega_j\omega_\ell\omega_r}{(\wosc+\omega_\ell +\omega_j)(\wosc+\omega_j+\omega_r)}\cos[(k_\ell -k_r)x] \, .
\end{equation}
\end{widetext}

\begin{figure}
\centering
\subfigure(a){\includegraphics[height=5.5cm, width=7cm]{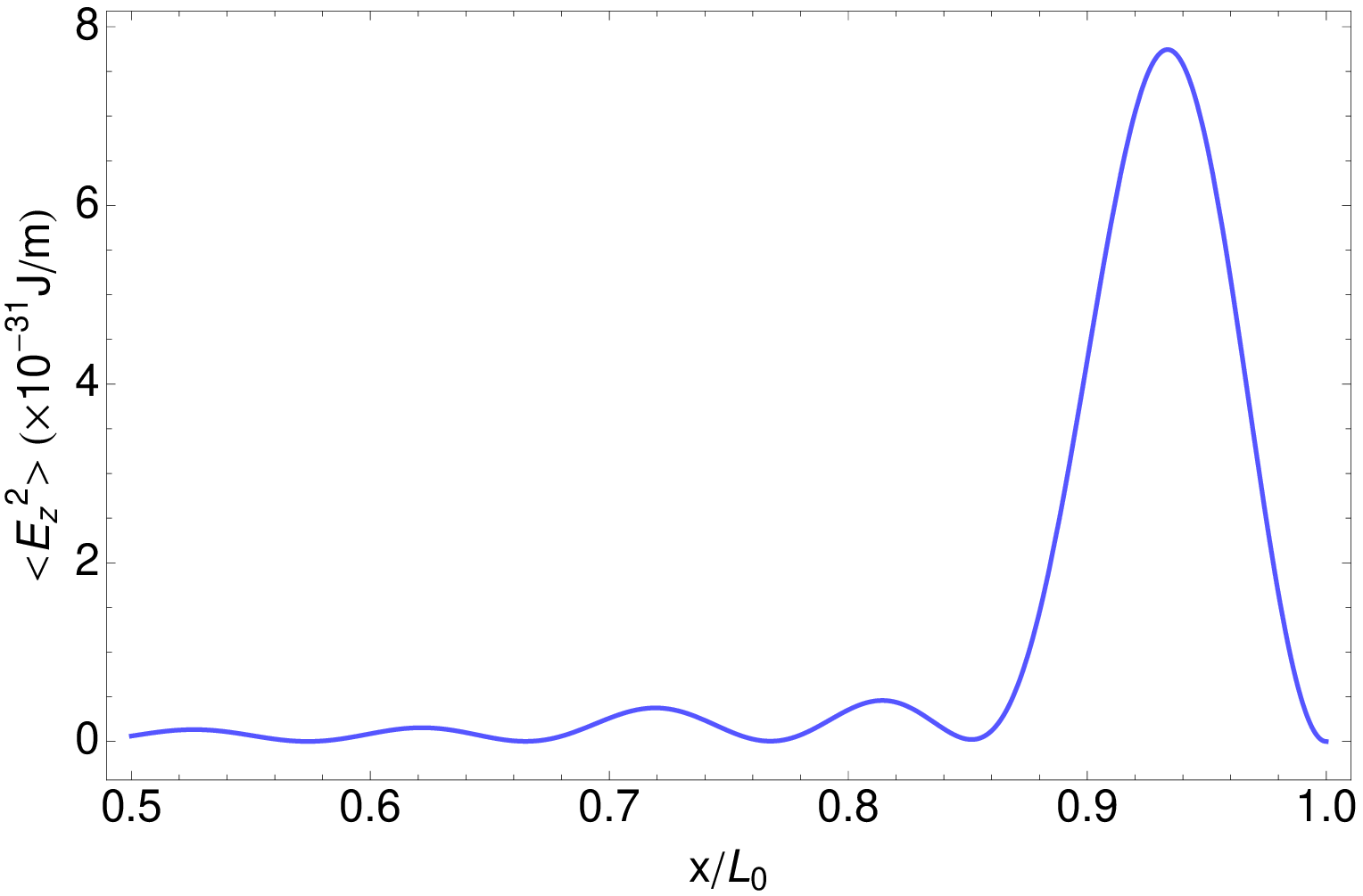}}
\subfigure(b){\includegraphics[height=5.5cm, width=7cm]{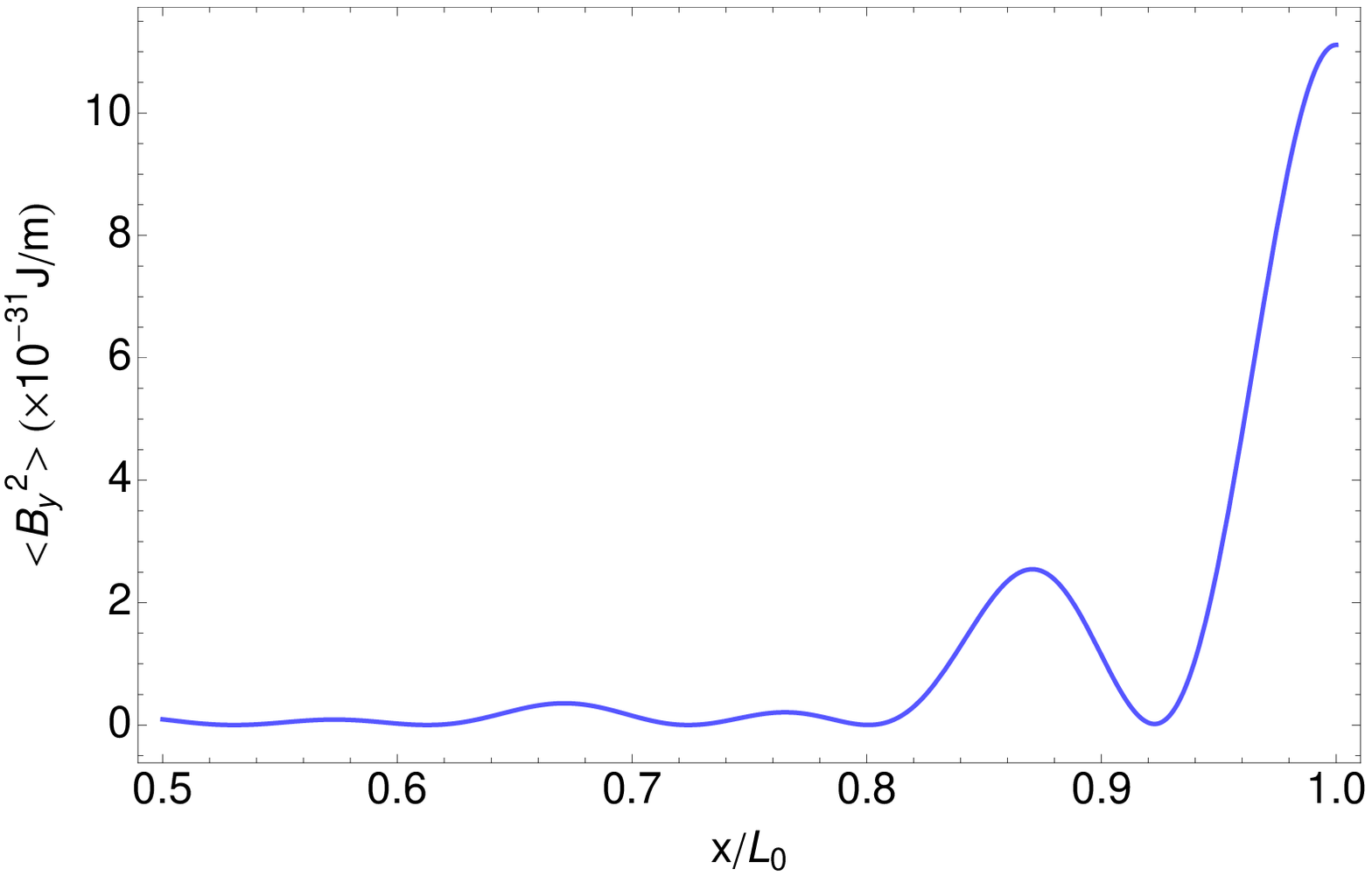}}
\subfigure(c){\includegraphics[height=5.5cm, width=7cm]{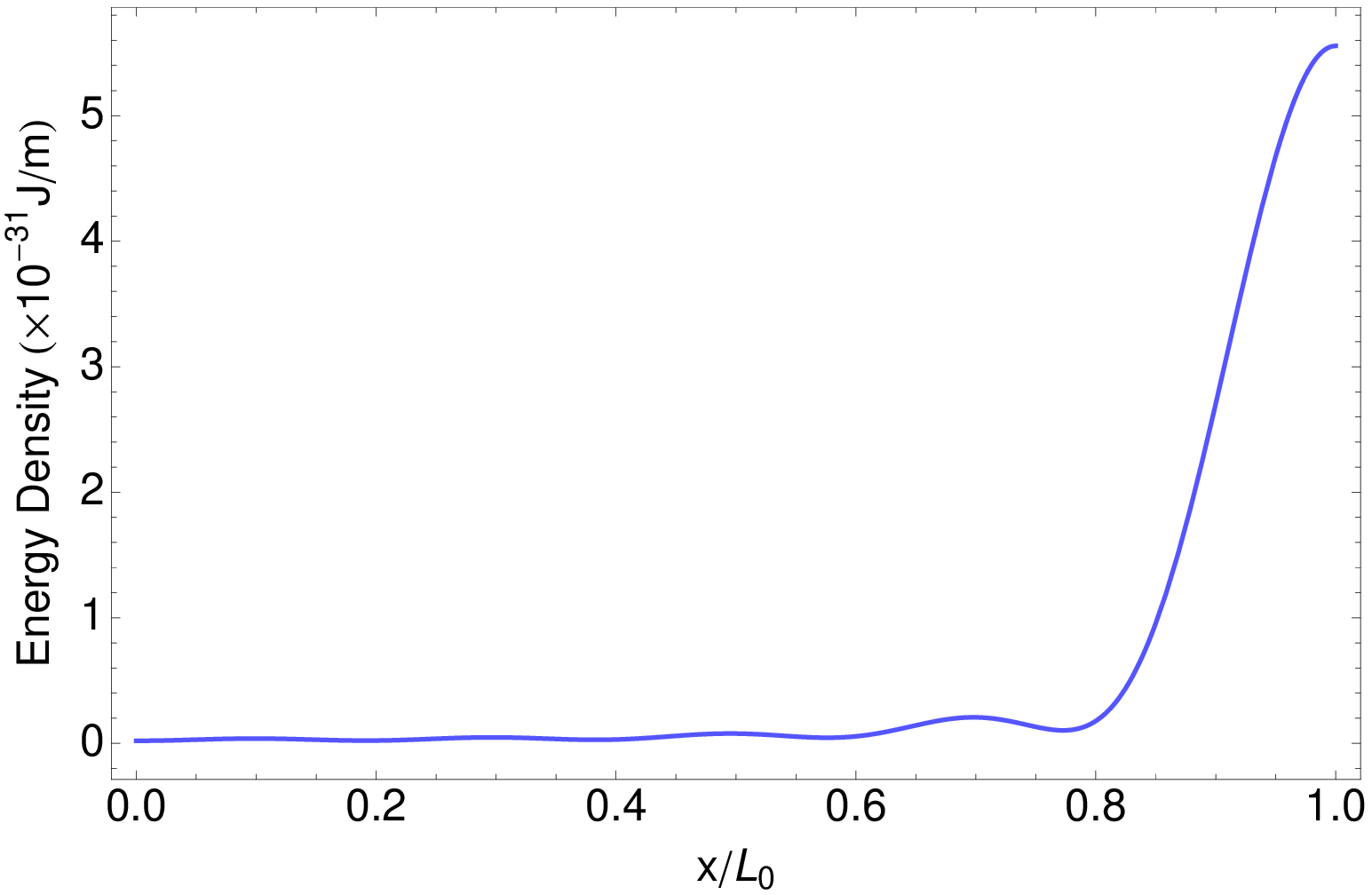}}
\caption{(color online) Plots (a) and (b) show the corrections to the renormalized electric and magnetic energy density, respectively, as a function of the position inside the cavity. The total field energy density is shown in plot (c). The numerical values used for the relevant parameters are: $L_0=10 \,\mu \text{m}$, $M=10^{-11} \, \text{kg}$, $\omega_{osc}=10^5 \text{$s^{-1}$}$, $\omega_{cut}=10^{15} \text{$s^{-1}$}$, which are typical values of a MEMS (microelectromechanical system).}
\label{graficoEB-1D}
\end{figure}

\begin{figure}
\centering
\includegraphics[height=5cm, width=7cm]{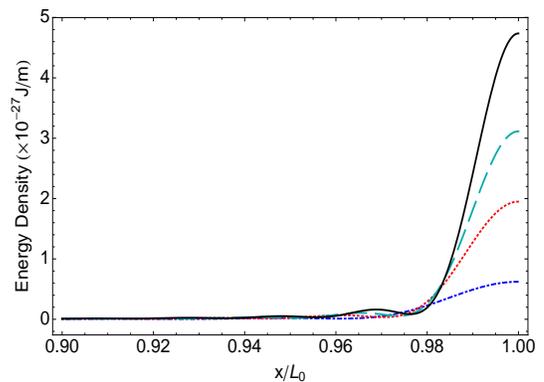}
\caption{(color online) Change of the renormalized electromagnetic energy density, compared to the static walls case, in the very proximity of the moving mirror, for different values of the cutoff frequency: $\omega_{cut}=6 \times10^{15}\, \text{s}^{-1}$ (blue dot-dashed line), $\omega_{cut}=8\times 10^{15} \, \text{s}^{-1}$ (red dotted line), $\omega_{cut}=9\times 10^{15} \, \text{s}^{-1}$ (green dashed line), $\omega_{cut}= 10^{16} \, \text{s}^{-1}$ (black continuous line). The  change of the energy density becomes more and more localized at the wall's equilibrium position with increasing cutoff frequency. The numerical values of the other parameters are $L_0=10 \, \mu m$, $M=10^{-11}\, \text{kg}$ and $\omega_{osc}= 10^5 \, \text{s}^{-1}$.}
\label{graficoED-1D}
\end{figure}

This quantity can be evaluated numerically. In Fig. $\ref{graficoEB-1D}$ we show the corrections to the electric and magnetic components of the electromagnetic energy density, as well as the correction to the total energy density, in the proximity of the mobile wall, where the effects we are investigating are more relevant. We have used the typical value of the
mass of a commercial MEMS, which is $M \simeq10^{-11}\, \text{kg}$ \cite{Gonzales}. However, much smaller masses in the range $10^{-15}-10^{-21}\, \text{kg}$ can be obtained nowadays in optomechanical devices
\cite{MG09,Schwab,Girvin}, and this should allow us to make even more significant the effect we are considering, because it scales as $1/M$.
The energy densities are plotted for typical values, $\omega_{osc}=10^5 \, \text{kg}$, $L_0=10 \, \mu m$ of the mirror's oscillation frequency and cavity length, respectively, and the cutoff frequency of a typical plasma frequency of a metal. The figures show that the motion of the wall significantly affects the field energy density inside the cavity and that this effect is particularly important near the moving wall. In addition, this effect becomes more and more relevant when the mass of the wall and its oscillation frequency are decreased, consistent with our previous results for the one-dimensional scalar field \cite{Pass-Butera}. In Fig. \ref{graficoED-1D}  we have plotted the correction to the vacuum field energy density caused by the wall's motion for different cutoff frequencies. We observe that when the cutoff frequency increases, the effect becomes more and more relevant and sharply localized in the vicinity of the wall's equilibrium position.

The changes to the energy density can be probed exploiting the Casimir-Polder dispersion interactions with an electrically or magnetically polarizable body placed inside the cavity. If this polarizable body has static electric polarizability $\alpha_E$ and static magnetic polarizability $\alpha_M$, its interaction energy with the electric and magnetic field fluctuations, under appropriate conditions, can be written as \cite{CP69,PPT98}
\begin{equation}
\delta E = - \frac 12 \alpha_E \langle E^2(x_{pb}) \rangle - \frac 12 \alpha_M \langle B^2(x_{pb}) \rangle \, ,
\label{Casimir-Polder}
\end{equation}
where $x_{pb}$ is the position of the polarizable body. Thus the interaction with the polarizable body permits to measure both electric and magnetic energy densities.

\section{\label{sec:level3}The three-dimensional scalar case}

We now discuss the case of a scalar field in a three-dimensional cavity. We consider a three-dimensional massless scalar field that satisfies Dirichlet boundary conditions $\phi(0,y,z)=\phi(L_x(t),y,z)=0$, inside a three-dimensional cavity with walls along the axis $x,\, y,\, z$ of length $L_x, \, L_y, L_z$, respectively. One of the two walls perpendicular to the $x$ axis is free to move and its position is $x=L_x(t)$, with $L_x(t)$ the time-dependent cavity length along the $x$ direction. All other boundaries are fixed in space. The movable wall has mass $M$ and, similarly to the case discussed in the previous section, it is bounded by a (harmonic) potential $V(q)$ at its equilibrium position.

In order to deal with our three-dimensional problem, we need to generalize the effective Hamiltonian \eqref{H0},\eqref{Hint-Giulio} used in the previous section to the three-dimensional case. We start with the situation in which all walls have fixed positions. In this case, the scalar field is given by
\begin{equation}\label{phi-3D}
  \phi(\textbf{r},t)=\sum_\textbf{n} \sqrt{\frac{\hbar c^2}{\omega_\textbf{n}SL_x}}\sin\left(q_x^\textbf{n} x\right)e^{i\textbf{q}_\|^\textbf{n} \cdot \textbf{r}_\|}e^{-i\omega_\textbf{n}t}+\text{H.c} \, ,
\end{equation}
where we have used periodic boundary conditions in the $y$ and $z$ directions, and
\begin{equation}\label{Def-k-w}\begin{split}
&\omega_\textbf{n}=c\sqrt{\left(\frac{n_x\pi}{L_x}\right)^2+\frac{(2\pi)^2}S(n_y^2+n_z^2)} \, ,\\
&\textbf{q}_\|^\textbf{n}=\frac{2\pi}{\sqrt{S}}(n_y\hat{y}+n_z\hat{z}), \, \, q_x^\textbf{n}= \frac{\pi}L_x n_x \, ,
\end{split}\end{equation}
are, respectively, the frequency and the wave vector components along the $yz$ plane and the $x$ direction. They depend on the three integer numbers $\textbf{n}=(n_x,\, n_y,\, n_z)$. We have assumed $L_y=L_z$ and defined $S=L_yL_z$; also, we have defined the component $\textbf{r}_\| = (y\hat{y}+z\hat{z})$ of the position in the plane $yz$.

When the wall perpendicular to the $x$ axis at $L_x(t)$ is movable, we can write the field in terms of a set of an instantaneous basis \cite{Mundarain-Neto,Crocce-Dalvit-Mazzitelli},
\begin{equation}
\phi (\textbf{r},t)=\sum_{\textbf{n}}a_\textbf{n} \phi_{\textbf{n}}(\textbf{r},t) + \text{H.c.} \, ,
\label{Field-operator}
\end{equation}
where the mode functions $\phi_{\textbf{n}}(\textbf{r},t)$ can be expanded in terms of a set of
modes $Q_\textbf{k}^{(\textbf{n})}$ which satisfy the massless Klein-Gordon equation with the instantaneous boundary condition at the movable wall's position,
\begin{equation}\label{phi-3D-modi}
  \phi_\textbf{n}(\textbf{r},t)=\sum_\textbf{k}Q^{(\textbf{n})}_\textbf{k}(t)\sqrt{\frac 1{\omega_{\textbf{k}}SL_x}}\sin\left(q_x^\textbf{k} x\right)
  e^{i\textbf{q}_\|^\textbf{k} \cdot \textbf{r}_\|} \,
\end{equation}
where $\textbf{k}=(k_x, \, k_y, \, k_z)$, with $k_x \, ,k_y \, ,k_z$ integer numbers, and the definitions \eqref{Def-k-w} have been used.

The field inside the cavity interacts with the movable wall through the radiation pressure. Because the wall is free to move along the $x$ direction only, we must first obtain the component of the radiation pressure along that axis. The relevant energy-momentum tensor component is thus \cite{Passante}
\begin{equation}\label{Tetaxx-3D}
  \Theta^{xx}=\frac{1}{2}\left[\left(\frac{\partial \phi}{\partial x}\right)^2+\frac 1{c^2}\left(\frac{\partial \phi}{\partial t}\right)^2-\left(\frac{\partial \phi}{\partial y}\right)^2-\left(\frac{\partial \phi}{\partial z}\right)^2\right] \, ,
\end{equation}
which involves both space and time derivatives of the field operator evaluated at the wall's position.
After calculating this quantity in the reference frame comoving with the movable wall, and then making a Lorentz transformation to the laboratory frame in the nonrelativistic limit \cite{Moore70}, we obtain that only the term $\frac{1}{2}\left(\frac{\partial \phi}{\partial x}\right)^2$ contributes to the force due to the radiation pressure, while the term involving the time derivative is negligible being proportional to $\dot{L}_x^2(t)$.

Therefore, the equations of motion for the field operator and the wall's position, in the nonrelativistic limit, are
\begin{equation}\label{eq-moto-3D}\begin{split}
&\Box\phi=0 \, ,\\
&m\ddot{q}=-\frac{\partial V(q)}{\partial q}+\frac{1}{2}\left(\frac{\partial \phi}{\partial x}\right)^2 \, ,
\end{split}\end{equation}
where $-\frac{\partial V(q)}{\partial q}$ is the force due to the potential binding the wall to its equilibrium position, and for simplicity the wall's position has been denoted by $q$.
Using Eq. $\eqref{phi-3D-modi}$ and the orthogonality of the mode functions, it is possible to show that the equations of motion $\eqref{eq-moto-3D}$ are equivalent to the following set of equations
\begin{widetext}
\begin{equation}\begin{split}\label{EQ-moto}
    &\ddot Q_\textbf{k}^{(\textbf{n})}+\omega_\textbf{k}^2(t)Q_\textbf{k}^{(\textbf{n})}=2\lambda(t)\sum_\textbf{j}g_{\textbf{kj}}\dot Q_\textbf{j}^{(\textbf{n})}+\dot\lambda(t)\sum_\textbf{j}g_{\textbf{kj}}Q_\textbf{j}^{(\textbf{n})}+\lambda^2(t)\sum_{\textbf{j,l}}g_{\textbf{jk}}g_{\textbf{jl}}Q_\textbf{l}^{(\textbf{n})} \, ,\\
    &m\ddot q=-\frac{\partial V(q)}{\partial q}+\frac{1}{q(t)}\sum_{\textbf{kjn}}(-1)^{k_x+j_x}Q_\textbf{k}^{(\textbf{n})} Q_\textbf{j}^{(\textbf{n})}\omega_{k_x}\omega_{j_x}\delta_{k_y,-j_y}\delta_{k_z,-j_z} \, ,
\end{split}\end{equation}
\end{widetext}
where $\textbf{j}=(j_x, \, j_y, \, j_z)$ indicates a set of three integer numbers, $\omega_{k_x}=\pi k_x/L_x$, $\omega_{j_x}=\pi j_x/L_x$, $\lambda={\dot{L}}_x(t)/{L_x(t)}$ and
\begin{equation}\label{gkj-3D}
  g_{\textbf{kj}}=\left\{
  \begin{array}{ll}
  (-1)^{k_x+j_x}\frac{2k_xj_x}{j_x^2-k_x^2}\,\delta_{k_y,-j_y}\delta_{k_z,-j_z} & (k_x\neq j_x)  \\
   0 & (k_x=j_x) \, .
   \end{array}
    \right.
\end{equation}
Following a procedure analogous to that used in Ref. \cite{LawField-Mirror}, we can show that Eqs. $\eqref{EQ-moto}$ can be obtained from a set of Euler-Lagrange equations relative to an appropriate Lagrangian. The corresponding Hamiltonian, associated with equations of motion $(\ref{EQ-moto})$, after a canonical quantization procedure and renormalization, is
\begin{equation}\label{H-Law-Gamma-3D}
  H=\frac{\left(p+\Gamma\right)^2}{2m}+V(q)+\hbar\sum_{\textbf{k}}\omega_{\textbf{k}}a^{\dag}_{\textbf{k}}a_{\textbf{k}}-E_{Cas} \, ,
\end{equation}
where $\Gamma$ is the following operator
\begin{equation}\label{Gamma-3D}
  \Gamma=\frac{i\hbar}{2q}\sum_{\textbf{kj}}g_{\textbf{kj}}\left(\frac{\wkG}{\wjG}\right)^{1/2}\left[\adkG\adjG-\akG\ajG+\adkG\ajG-\adjG\akG\right] \, ,
\end{equation}
and the quantity $E_{Cas}$ is the Casimir energy for the two-fixed-walls configuration.

Our Hamiltonian \eqref{H-Law-Gamma-3D} is the generalization of the Hamiltonian obtained by Law in \cite{LawField-Mirror} to the three-dimensional case. It shows a nonlinear character of the coupling between the field and the movable wall; however, in many situations the wall is bounded to an equilibrium position by an external potential, and the effects of the radiation pressure can be treated as a small perturbation. We assume this is the case in our system, and thus we now derive a linearized form of the Hamiltonian $(\ref{H-Law-Gamma-3D})$.
We suppose that the mobile wall is confined near the equilibrium position $L_0$ by the potential $V(q)$ . When the displacement of the wall $x_m=q-L_0$ is small compared to $L_0$, we can write
\begin{equation}\label{App-lineare-3D}\begin{split}
  &\Gamma\approx \left. \Gamma \right|_{q=L_0} \equiv \Gamma_0 \, ,\\
  &a_\textbf{k}(q)\approx\akGl+\frac{x_m}{2\wkGl}\left(\frac{\partial\wkG}{\partial q}\right)_{q=L_0}\adkGl \, ,\\
  &\wkG(q)\approx\wkGl+x_m\left(\frac{\partial\wkG}{\partial q}\right)_{q=L_0}
\end{split}\end{equation}
where $\akGl$ and $\wkGl$ are, respectively, the annihilation operator and the frequency associated with the equilibrium position $L_0$.
We now substitute Eqs. $\eqref{App-lineare-3D}$ into the Hamiltonian $\eqref{H-Law-Gamma-3D}$ and make a unitary transformation $H'=T^{\dag}HT$, where the transformation operator is given by
\begin{equation}
  T=\exp\{ix_m\Gamma_0/\hbar\} \, .
\end{equation}
We also assume that $V(q)$ is a harmonic potential with frequency $\wosc$. After lengthy algebraic calculations,  we can write our Hamiltonian in the form $H=H_0+H_{int}$, where
\begin{equation}\label{H0-3D}
  H_0=\hbar\wosc b^{\dag}b+\hbar\sum_{\textbf{k}}\wkGl\adkGl\akGl
\end{equation}
is an unperturbed Hamiltonian where $b$ and $b^{\dag}$ are the annihilation and creation operators of the movable wall, respectively; the effective Hamiltonian $H_{int}$ describing the interaction between the three-dimensional massless scalar field and the movable wall is
\begin{equation}\label{Hint-3D}
  H_{int}=\sum_{\textbf{k,j}}\CkjG\left\{(b+b^{\dag})\mathcal{N}[(\akGl+\adkGl)(\ajGl+\adjGl)]\right\} \, ,
\end{equation}
where
\begin{equation}\label{Ckj-3D}
  \CkjG=\frac{\hbar}{2}\sqrt{\frac{\hbar}{2M\wosc}}\left[\frac{\partial\wkGl}{\partial q}\Bigg|_{L_0}\delta_{\textbf{k,j}}-\frac{g_{\textbf{k,j}}}{L_0}\left(\frac{\wkGl}{\wjGl}\right)^{1/2}\wkGl\right] \, .
\end{equation}
The quantity $ \CkjG$ given in \eqref{Ckj-3D} is the new coupling constant for the wall-field interaction in the three-dimensional case, which can be compared with the analogous one-dimensional coupling constant of  Sec. \ref{sec:level2}, given by Eq. \eqref{Ckj} .

From our effective Hamiltonian \eqref{Hint-3D}, using first-order perturbation theory we can obtain the interacting (dressed) ground state of the field-mirror system, having virtual excitations of both field and mirror (we are using, for the states, the same notations as in  Sec. \ref{sec:level2}),
\begin{equation}\label{ground-3D}
  |g\rangle=|\{0_{\textbf{p}}\},0\rangle+\sum_{\textbf{k} \textbf{j}}\DkjG|\{1_{\textbf{k}},1_{\textbf{j}}\},1\rangle \, ,
\end{equation}
where we have defined
\begin{equation}\label{DkjG}
   \DkjG=\frac{\CkjG}{\hbar(\wosc+\wkG+\wjG)} \, .
\end{equation}

We can now evaluate the average number of excitations of the three-dimensional cavity in the field mode specified by the integers
$\textbf{m}= (m_x, \, m_y, \, m_z)$,
in the dressed ground state $(\ref{ground-3D})$, due to the quantum fluctuations of the movable wall's position. It results in
\begin{widetext}
\begin{equation}\label{Nm-tot-3D}\begin{split}
&\langle g|N_{\textbf{m}}|g\rangle\equiv\langle g|N_{(m_x,m_y,m_z)}|g\rangle =\frac{\pi^4 \hbar c^4}{2ML_0^6}\frac{m_x^4}{\wosc (\omega_\textbf{m})^2(\wosc+2\omega_\textbf{m})^2}\\
&+\frac{\hbar}{2ML_0^2}\sum_{m_x'\neq m_x} \frac{(m_x m_x')^2}{((m_x')^2-m_x^2)^2}\frac{(\omega_\textbf{m}^2-\omega_{(m_x',m_y,m_z)}^2)^2}{\wosc\omega_\textbf{m}\omega_{(m_x',m_y,m_z)}
(\wosc+\omega_\textbf{m}+\omega_{(m_x',m_y,m_z)})^2} \, ,
\end{split}\end{equation}
\end{widetext}
where we have used Eq. $(\ref{Def-k-w})$.

This is the photon spectrum due to the motion of the mirror. The field excitations originate from the vacuum state as a consequence of the mirror-field interaction.

The one-dimensional case can be recovered from our result $\eqref{Nm-tot-3D}$ by setting $m_y=m_z=0$ and obtaining
\begin{equation}\label{N-1D} \begin{split}
&\langle g|N_{(m_x,0,0)}|g\rangle \\
&=\frac {\hbar c^4}{2ML_0^2\wosc}\sum_{m_x'}\frac{\omega_{(m_x,0,0)}\omega_{(m_x',0,0)}}{(\wosc+\omega_{(m_x,0,0)}+\omega_{(m_x',0,0)})^2} \, ,
\end{split}\end{equation}
which coincides with the result already obtained in \cite{Pass-Butera} for the one-dimensional scalar case.

Equation \eqref{Nm-tot-3D} shows that, similarly to the one-dimensional case \cite{Pass-Butera}, the number of virtual quanta inside the cavity decreases with increasing oscillation frequency $\wosc$ and mass $M$ of the mobile wall. From a physical point of view, this is due to the fact that when the mirror oscillation frequency increases, its action in mediating the effective wall-field interaction is weaker. An analogous consideration holds for the dependence from the mirror's mass.

In order to obtain local field quantities for the three-dimensional scalar field, which, as mentioned in the Introduction, are also useful to evaluate analogous quantities for the three-dimensional electromagnetic field, we first calculate the renormalized field propagator.
The renormalized propagator is the difference between two different Green's functions. The first function is the propagator of the field with Dirichlet boundary conditions on the dressed vacuum state $|g\rangle$ given by \eqref{ground-3D}, while the second one is the free-field propagator,
\begin{widetext}
\begin{equation}\label{phi-3D-rin}
 G_R(\bx ,t;\bxp ,t')=\langle g|\phi_{BC}(\textbf{x},t)\phi_{BC}(\textbf{x}',t')|g\rangle-\langle\{0_\textbf{r}\}|\phi_{un}(\textbf{x},t)\phi_{un}(\textbf{x}',t')|\{0_\textbf{r}\}\rangle=G_{R0}(\bx ,t;\bxp ,t')+\Delta G_R(\bx ,t;\bxp ,t') \, ,
\end{equation}
\end{widetext}
where $(\textbf{x},t)$ is the space-time coordinate, and subscripts $BC$ and $un$ indicate quantities in the presence of boundary conditions and in the unbounded space, respectively.

The renormalized Green's function of the system is
\begin{equation}
\label{Green3D}
 G_R(\bx ,t;\bxp ,t')=G_{R0}(\bx ,t;\bxp ,t')+\Delta G_R(\bx ,t;\bxp ,t') \, ,
 \end{equation}
where $G_{R0}(\bx ,t;\bxp ,t')$ is a zeroth-order term in the effective field-wall interaction, giving the renormalized propagator (meaning that the unbounded-field propagator has been subtracted) for a fixed-walls cavity, and $\Delta G_R(\bx ,t;\bxp ,t')$ is a first-order correction due to the motion of the mobile wall. An explicit calculation of these two terms yields

\begin{widetext}
\begin{equation}\label{Definizioni}\begin{split}
&G_{R0}(\bx ,t;\bxp ,t')=\langle\{0_\textbf{r}\}|\phi_{BC}(\textbf{x},t)\phi_{BC}(\textbf{x}',t')|\{0_\textbf{r}\}\rangle-\langle\{0_\textbf{r}\}|\phi_{un}(\textbf{x},t)\phi_{un}(\textbf{x}',t')|\{0_\textbf{r}\}\rangle\\
&=\Bigg[\sum_\mathbf{m}\left(\frac{\hbar c^2}{\omega_\mathbf{m} SL_0}\right)\sin\left(\frac{m_x\pi}{L_0}x\right)\sin\left(\frac{m_x\pi}{L_0}x'\right) e^{i\textbf{k}^\textbf{m}_{\|}\cdot(\textbf{r}_{\|}-\textbf{r}'_{\|})}e^{-i\omega_\textbf{m}(t-t')}-\int\frac{d\textbf{p}}{(2\pi)^3}\frac{\hbar c^2}{2\wpG}e^{-i\wpG(t-t')}e^{ik_{\textbf{p}}\cdot(\textbf{x}-\textbf{x}')} \Bigg] \, ,\\
&\Delta G_R(\bx ,t;\bxp ,t')=\sum_{\textbf{mj}}\sum_{\textbf{r}}4 D_{\textbf{mj}} D_{\textbf{jr}}\left[ u_\textbf{m}(\bx ,t)\urG^*(\bxp ,t')+u_\textbf{m}^*(\bx ,t)\urG(\bxp ,t')\right] \, ,
\end{split}\end{equation}
where we have defined
\begin{equation}\label{Un}
  u_{\textbf{m}}(\textbf{x},t)\equiv u_\textbf{m}(x,\textbf{r}_\|,t)=\sqrt\frac{\hbar c^2}{\omega_{\textbf{m}} SL_0}\sin\left(\frac{m_x\pi}{L_0}x\right)e^{i\textbf{k}^\textbf{m}_\|\cdot \textbf{r}_\|}e^{-i\omega_\textbf{m} t} \, .
\end{equation}
\end{widetext}
The zeroth-order term takes into account the fact that the field is confined, while the correction term also takes into account the field-wall interaction and the wall's quantum position fluctuations.

From the expressions obtained above for the scalar propagator, we can now obtain the energy density of the scalar field in the cavity by applying the appropriate differential operators according to Eq.
$(\ref{Teta-vuoto-BC})$,
\begin{equation}\label{ro-3D}\begin{split}
  \langle\rho\rangle&=\frac{1}{2}\lim_{(\bxp ,t')\rightarrow (\bx ,t)}\left(c^{-2} \partial_t\partial_{t'}+\nabla_{\textbf{x}}\cdot\nabla_{\textbf{x}'}\right)G_R(\bx ,t;\bxp ,t')\\
 &=\frac{1}{2}\lim_{(\bxp ,t')\rightarrow (\bx ,t)}\left(c^{-2} \partial_t\partial_{t'}+\nabla_{\textbf{x}}\cdot\nabla_{\textbf{x}'}\right)[G_{R0}+\Delta G_R] \\
 &=\langle\rho_0\rangle + \langle\Delta\rho\rangle \, .\\
 &
\end{split}\end{equation}

Therefore, the field energy density can also be written as a zero-order term $\langle\rho_0\rangle$ plus a first-order correction $\langle\Delta\rho\rangle$ related to the quantum fluctuations of the wall's position and radiation pressure, respectively, given by
\begin{widetext}
\begin{equation}\label{Conti}\begin{split}
\langle\rho_0(\mathbf{x})\rangle&=\frac{1}{2}\lim_{(\bxp ,t')\rightarrow (\bx ,t)}
\left(c^{-2}\partial_t\partial_{t'}+\nabla_{\textbf{x}}\cdot\nabla_{\textbf{x}'}\right)G_{R0}(\bx ,t;\bxp ,t')=-\frac{\pi^2\hbar c}{1440L_0^4}-\sum_{\textbf{p}}\frac{\hbar}{SL_0}\cos\left(\frac{2p_x\pi}{L_0}x\right)\frac{c^2(p_x^2+2p_{\|}^2)}{2\wpG}e^{-\eta\wpG} \, ,\\
\langle\Delta\rho(\mathbf{x})\rangle&=\frac{1}{2}\lim_{(\bxp ,t')\rightarrow (\bx ,t)}\left(c^{-2}\partial_t\partial_{t'}+\nabla_{\textbf{x}}\cdot\nabla_{\textbf{x}'}\right)
\Delta G_R(\bx ,t;\bxp ,t')\\
&=\frac{1}{2}\Bigg[\sum_{\textbf{m}\textbf{j}\textbf{r}}8D_{\textbf{mj}}D_{\textbf{jr}}\frac{\hbar c^2}{SL_0}\Bigg[\left(\frac{\sqrt{\wrG\omega_\mathbf{m}}}{c^2}+ \frac{4\pi^2}S
\frac{m_z r_z+m_yr_y}{\sqrt{\wrG\omega_\textbf{m}}}\right)\sin\left(\frac{m_x\pi}{L_0}x\right)\sin\left(\frac{r_x\pi}{L_0}x\right)\\
&+\frac{m_x r_x}{\sqrt{\wrG\omega_\mathbf{m}}}\left(\frac{\pi}{L_0}\right)^2\cos\left(\frac{m_x\pi}{L_0}x\right)
\cos\left(\frac{r_x\pi}{L_0}x\right)\Bigg]\cos[(\textbf{k}_{\|}^\mathbf{m}-\textbf{k}_{\|}^{\mathbf{r}})\cdot \textbf{r}_{\|}] \Bigg] \, ,
\end{split}\end{equation}
\end{widetext}
where $\eta^{-1}$ is an ultraviolet cutoff parameter related to the plasma frequency of the cavity walls. We note that in Eq. $\eqref{Conti}$, because of the relations $\eqref{Ckj-3D}$ and $\eqref{DkjG}$, we obtain $\textbf{k}_{\|}^\mathbf{m}=\textbf{k}_{\|}^{\mathbf{r}}$ and thus the correction to the energy density does not depend on $\textbf{r}_{\|}$.

The zero-order term contains the well-known Casimir energy density
$-\frac{\pi^2\hbar c}{1440L_0^4}$ for a three-dimensional scalar field with Dirichlet boundary conditions (see \cite{Ford-Svaiter,Milton1,Milton11}, for example), and a term dependent on the position inside the cavity that, in the limit of perfectly conducting walls (infinite cutoff frequency, that is $\eta\rightarrow 0$, or $\omega_c\rightarrow\infty$) diverges in the proximity of the wall's positions. This divergence is related to the presence of a sharp ideal boundary condition, in agreement with the results in Refs. \cite{Ford-Svaiter, Milton1, Milton11}, and in Ref. \cite{BP12} for a single wall. By integration over $x$, the two contributions in $\langle\rho_0\rangle$ yield the three-dimensional scalar field Casimir energy $E_0=-\frac{\pi^2\hbar c}{1440L_0^3}$ between two fixed walls.

\begin{figure}
\centering
\includegraphics[height=5cm, width=7cm]{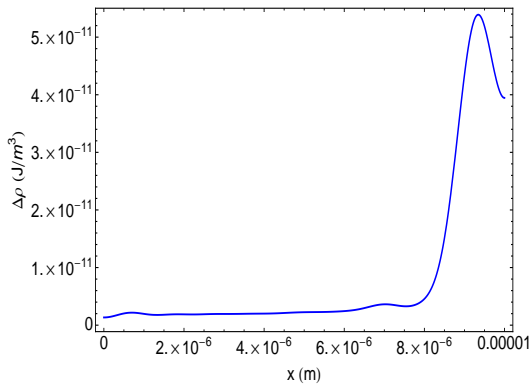}
\caption{(color online) Correction to the scalar energy density of the field inside the three-dimensional cavity with respect to the fixed-wall configuration. The numerical values of the parameters used are: $L_0=10 \, \text{$\mu m$}$, $L_y=L_z=0.5 \times 10^{-4}\, \text{m}$, $M=10^{-11} \, \text{kg}$, $\omega_{osc}=10^5 \, \text{$s^{-1}$}$ e $\omega_{cut}=10^{15} \, \text{$s^{-1}$}$.}\label{grafico3D-scalare}
\end{figure}

The correction $\langle\Delta\rho\rangle$ obtained in Eq. $(\ref{Conti})$ gives the change we are looking for of the field energy density due to the wall's movement. It can be evaluated numerically and it is plotted in Fig. \ref{grafico3D-scalare} with parameters such that $L_y=L_z \gg L_0$. Our results show that, in the three-dimensional scalar case as well, the energy density in a cavity with a movable wall differs from that of a fixed wall configuration. The change of the renormalized field energy density and of the renormalized field fluctuations is particularly significant in the very proximity of the movable wall. This new effect can be experimentally probed using the Casimir-Polder interaction with a polarizable body placed inside the cavity, analogously to the electromagnetic one-dimensional case discussed in the previous section.
This effect is more relevant for decreasing mass and oscillation frequency of the mobile wall. Because in actual optomechanical experiments it is possible to obtain extremely small masses, of the order of $10^{-21}\, \text{kg}$, corrections to Casimir-Polder potentials of some few percent seem realistic if the polarizable body is placed very close to the movable wall. These effects should be currently measurable since nowadays Casimir interactions can be measured with high precision \cite{Buhmann12,Lamoreaux05}.

Our case of a three-dimensional scalar field has some qualitatively significant difference compared to the scalar one-dimensional case discussed in \cite{Pass-Butera}.  In fact, contrary to the one-dimensional case, where the maximum change of the energy density is at the wall's equilibrium position, in the present three-dimensional case the peak is shifted with respect to the equilibrium position of the wall, as Fig. \ref{grafico3D-scalare} clearly shows. The distance of the peak from the equilibrium position is strongly related to the cutoff frequency of the cavity walls. A numerical evaluation carried out for different values of the cutoff frequency shows that when increasing the cutoff frequency, the peak approaches more and more the wall's equilibrium position. Finally, we wish to mention that our results for the three-dimensional massless scalar field can be the basis for obtaining the field energy densities for the three-dimensional electromagnetic field too, both for transverse electric and magnetic modes, using the relations outlined in Sec. \ref{sec:local-formalism}.

\section{\label{sec:Conclusion}Conclusions}

We have considered the interaction between a moving conducting wall, whose mechanical degrees of freedom are treated quantum mechanically, and a field. We have considered the cases of the electromagnetic field in a one-dimensional cavity and a massless scalar field in a three-dimensional cavity, generalizing previous results obtained for the simpler case of a one-dimensional scalar field. The movement of the wall, which we have assumed bound to its equilibrium position by a harmonic potential, yields an effective wall-field interaction and an effective interaction between the field modes, mediated by the mobile wall. For the one-dimensional electromagnetic case, using the Green's function formalism, we have been able to obtain the electric and magnetic energy densities, exploiting previous results obtained for the scalar one-dimensional case using the Law's effective Hamiltonian. For the three-dimensional scalar case, we have first generalized the Law's Hamiltonian, originally obtained only for the one-dimensional case, to our three-dimensional case with Dirichlet boundary conditions, and then obtained the renormalized Green's function in the interacting ground state. For both cases, we have evaluated the corrections to the field energy densities inside the cavity in the dressed ground state, and found that they are particularly significant in the vicinity of the movable wall. We have also found that these effects become more relevant with decreasing mass and oscillation frequency of the movable wall around its equilibrium position. We have also discussed measurability of these effects, exploiting Casimir-Polder interactions with a polarizable body placed inside the cavity, and shown that they should be observable by the optomechanical techniques available nowadays. Finally, we point out that our results for the three-dimensional scalar case could also be the basis for obtaining the renormalized electromagnetic energy densities for the transverse electric modes in a three-dimensional cavity with a mobile wall; obtaining the transverse magnetic modes would require solving an analogous scalar problem with von Neumann boundary conditions. We shall discuss these points in a subsequent paper.

\section*{ACKNOWLEDGMENTS}
The authors wish to thank Giulio Butera, Lucia Rizzuto and Salvatore Spagnolo for many discussions on the subject of this paper. Financial support by the Julian Schwinger Foundation, MIUR and CRRNSM is gratefully acknowledged.

\end{document}